\documentclass[aps,prb,showpacs,twocolumn,superscriptaddress,10pt]{revtex4-1}

\usepackage{amsmath,amssymb,amsfonts,amsthm}
\usepackage[utf8]{inputenc}
\usepackage{graphicx}
\usepackage[caption=false]{subfig}
\usepackage[pdftex,bookmarks=false,colorlinks=true,linkcolor=blue,
citecolor=blue,filecolor=black,urlcolor=blue]{hyperref}

\providecommand{\R}{{\mathbb{R}}}

\providecommand{\ud}{\,\mathrm{d}}
\providecommand{\abs}[1]{\lvert#1\rvert}
\providecommand{\averageop}[3]{\left\langle#1\left\lvert#2\right\rvert#3\right\rangle}

\providecommand{\xc}{\mathrm{xc}}
\providecommand{\ext}{\mathrm{ext}}
\providecommand{\SCE}{\mathrm{SCE}}

\providecommand{\bvec}[1]{\mathbf{#1}}
\providecommand{\vf}{\bvec{f}}
\providecommand{\vs}{\bvec{s}}
\renewcommand{\vr}{\bvec{r}}

\begin{document}

\title{Towards the Kantorovich dual solution for\\
strictly correlated electrons in atoms and molecules}
\affiliation{Mathematics Department,
Technische Universit\"at M\"unchen,
Boltzmannstra{\ss}e 3,
85748 Garching bei M\"unchen, Germany}
\affiliation{Computational Research Division,
Lawrence Berkeley National Laboratory,
Berkeley, California 94720, USA}
\author{Christian~B.~Mendl}
\affiliation{Mathematics Department,
Technische Universit\"at M\"unchen,
Boltzmannstra{\ss}e 3,
85748 Garching bei M\"unchen, Germany}
\author{Lin~Lin}
\affiliation{Computational Research Division,
Lawrence Berkeley National Laboratory,
Berkeley, California 94720, USA}


\begin{abstract}
	The many-body Coulomb repulsive energy of strictly correlated
	electrons provides direct information of the exact Hohenberg-Kohn
	exchange-correlation functional in the strong interaction limit.
	Until now the treatment of strictly correlated electrons is based on
	the calculation of co-motion functions with the help of semi-analytic
	formulations. This procedure is system specific and has been limited
	to spherically symmetric atoms and strictly 1D systems. We develop a
	nested optimization method which solves the Kantorovich dual
	problem directly, and thus facilitates a general treatment of strictly correlated
	electrons for systems including atoms and small molecules.  
\end{abstract}

\maketitle

\section{Introduction}\label{sec:intro}

Kohn-Sham density functional theory (KSDFT) is the most widely used
electronic structure theory for systems with many electrons, from
gas phase molecules to condensed matter systems in
particular~\cite{HohenbergKohn1964,KohnSham1965}. It is in
principle an exact theory, and is able to  yield the exact
ground state energy and density using a fictitious system of
non-interacting electrons.  The key component of KSDFT is the
exchange-correlation functional.  Tremendous
progress has been made in the past three decades for constructing
approximate exchange-correlation functionals based on the known
information from the uniform electron
gas~\cite{PerdewZunger1981,Becke1988,LeeYangParr1988,PerdewBurkeErnzerhof1996}. However, such
approximate exchange-correlation functionals are known to fail
for strongly correlated systems, such as the chromium
dimer~\cite{AnderssonRoosMalmqvistEtAl1994} or Mott-Hubbard
insulators~\cite{LiechtensteinAnisimovZaanen1995,CohenSanchezYang2012}.
Several recent studies indicate that the construction of
exchange-correlation functionals for general strongly correlated systems
can be extremely
difficult~\cite{SanchezCohenYang2009,CohenSanchezYang2012,StoudenmireWagnerWhiteEtAl2012}.  

Recently, the behavior of the exchange-correlation functional has been
revealed in the limit of strictly correlated electrons
(SCE)~\cite{Seidl2007,OTformulation2012,MaletGiorgi2012,CFK2012}.  The
many-body Coulomb repulsive energy of SCE determines the \textit{exact}
exchange-correlation functional in the strong interaction limit, without
artificially breaking any symmetry of the system or introducing any
tunable parameters.  The information provided by the SCE limit is
complementary to that provided by the Kohn-Sham non-interacting kinetic energy. 
For given electron density profile, the SCE limit is described by minimizing
the many-body Coulomb interaction energy with respect to all
%antisymmetric 
wavefunctions in a $3 N$ dimensional space, where $N$ is
the number of electrons in the system, under the additional constraint
that the wavefunction is consistent with the electron density~\cite{Seidl2007}.
Mathematically, this daunting minimization task is an optimal transport
problem with Coulomb cost
function~\cite{DFTStrongInteracting2009,OTformulation2012,CFK2012}.
The optimal transport theory finds the optimal way for transferring
masses from one position to another. The theory dates back to Monge in
1781, and was significantly generalized by Kantorovich in
1942~\cite{Kantorovich1942,Kantorovich2006}.  We refer the readers to
Ref.~\onlinecite{Villani2008} for more detailed information.
From physical intuition, the optimal transport problem with Coulomb cost function can be simplified by
introducing $N$ \emph{co-motion functions} $f_i: \R^3 \to \R^3$~\cite{Seidl2007}.
These co-motion functions characterize the
relative positions of all the electrons with respect to one given
electron in the SCE limit. To the extent of our knowledge, in practice
the co-motion functions can only be determined for one
dimensional systems~\cite{MaletGiorgi2012} and spherically symmetric
atoms~\cite{Seidl2007,OTformulation2012,CFK2012}, with the help of
semi-analytic methods. Little is known about the shape or even the
existence of the co-motion functions for general systems including small
molecules. On the other hand, the optimal transport problem with Coulomb
cost function can be
equivalently solved by its dual formulation, called the Kantorovich dual
problem~\cite{Kantorovich1942,Kantorovich2006,Villani2008,OTformulation2012,GangboSwiech1998}.
The main advantage of the
Kantorovich dual problem is its potential applicability to general
systems, ranging from atoms and molecules to condensed matter systems.
However, the Kantorovich dual problem is formulated as a maximization
problem with an \textit{infinite} number of constraints, which is
impossible to be implemented directly. These limitations severely
restrict the applicability of the SCE limit to systems of practical
interest. In this paper, we develop a novel method that solves the
Kantorovich dual problem directly. We overcome the difficulty of
infinite number of constraints via a nested optimization approach. Our
method provides a more general treatment of the exchange-correlation
functional in the SCE limit for atoms and small molecules.

The rest of the paper is organized as follows: in
Section~\ref{sec:theory}, we briefly review the SCE limit, the optimal
transport formulation and the Kantorovich dual problem, and present the
nested optimization method for solving the Kantorovich dual problem
directly. In Section~\ref{sec:numerical}, we establish the
applicability and accuracy of this method for the 3D beryllium atom and
a model quantum wire system in 1D, for which accurate results can be
obtained semi-analytically using the co-motion formulation.
Next, we demonstrate the applicability of our method to a model trimer
with various number of electrons in 3D, for which the SCE limit cannot
be calculated by existing techniques. The conclusion and future work
is given in Section~\ref{sec:conclusion}.

\section{Theory}\label{sec:theory}

According to the Hohenberg-Kohn theorem~\cite{HohenbergKohn1964}, the ground state energy of
a system can be obtained by minimizing the following functional with respect to
the electron density $\rho(\vr)$:
\begin{equation}
	E[\rho] = F[\rho] + \int v_{\ext}(\vr) \rho(\vr) \ud\vr. 
	\label{eqn:HKfunctional}
\end{equation}
Here $v_{\ext}(\vr)$ is the external potential, and $F[\rho]$ is the
internal energy functional, which is a
\emph{universal functional} of the electron density and consists of
the kinetic energy and the Coulomb repulsive energy between the
electrons. Formally $F[\rho]$ is defined by minimizing over all the
antisymmetric wavefunctions $\Psi$ which are consistent with
$\rho(\vr)$ as
\begin{equation}
	F[\rho] = \min_{\Psi \to \rho} 
	\averageop{\Psi}{\hat{T}+\hat{V}_{ee}}{\Psi}.
	\label{eqn:Ffunctional}
\end{equation}
Here $\hat{T}=-\sum_{i=1}^{N}\frac{1}{2}\Delta_{i}$ is the kinetic
energy operator and $\hat{V}_{ee} = \sum_{i=1}^{N} \sum_{j>i}^{N}
\abs{\vr_i - \vr_j}^{-1}$ is the Coulomb repulsive energy operator.
$\Delta_{i}$ is the Laplacian operator on the $i$-th electron.
The strong interaction limit considers the situation when the Coulomb
repulsive energy dominates over the kinetic energy, in which case the
internal energy functional can be approximated
as~\cite{MaletGiorgi2012}
\begin{equation}
	\begin{split}
	F[\rho] &\approx \min_{\Psi \to \rho} 
	\averageop{\Psi}{\hat{T}}{\Psi} +
	\min_{\Psi \to \rho} 
	\averageop{\Psi}{\hat{V}_{ee}}{\Psi}\\
	& \equiv T_{s}[\rho] + V_{ee}^{\SCE}[\rho].
	\end{split}
	\label{eqn:SCEsplit}
\end{equation}
The first term $T_{s}[\rho]$ is the Kohn-Sham (KS) kinetic energy
functional corresponding to a non-interacting independent particle
system~\cite{KohnSham1965}.  The second term $V_{ee}^{\SCE}[\rho]$ is the minimal Coulomb
repulsive energy among all wavefunctions which are
consistent with $\rho(\vr)$, and the corresponding minimizer
characterizes the state of
``strictly correlated electrons'' (SCE). In terms of KSDFT, $V_{ee}^{\SCE}[\rho]$ is the
sum of the Hartree energy and the exchange correlation energy, and
the exchange-correlation functional in the SCE limit can be
recovered by
\begin{equation}
	E_{\xc}^{\SCE}[\rho] = V_{ee}^{\SCE}[\rho] - \frac12
	\iint \frac{\rho(\vr)\rho(\vr')}{\abs{\vr-\vr'}} \ud\vr \ud\vr'.
	\label{eqn:XCfunctional}
\end{equation}

Eq.~\eqref{eqn:SCEsplit} allows for treating the
kinetic energy functional and the Coulomb repulsive energy functional
on the same footing but with different numerical techniques. The minimization of 
the kinetic energy functional $T_{s}[\rho]$ gives rise to a energy minimization
problem or a nonlinear eigenvalue problem known as the Kohn-Sham
equations. Their efficient treatment has been extensively
explored in the past few decades. The minimization
of the SCE functional $V_{ee}^{\SCE}[\rho]$ gives rise to an optimal
transport problem, for which numerical methods are still very sparse.
From the mathematical point of view, the results for the optimal
transport problem are primarily concerned with quadratic cost functions. 
Rigorous treatment of the Coulomb cost function only appeared
recently~\cite{CFK2012} for $N = 2$,
and the formal description for general $N$ as well as results concerning the
dual formulation have been introduced
in Ref.~\onlinecite{OTformulation2012}. However, a proper mathematical
generalization concerning the existence and uniqueness of the co-motion functions
has not been achieved yet.

Formally, the optimal transport problem is solved by minimizing over all
$3 N$-dimensional wavefunctions that are consistent with
the given electron density $\rho(\vr)$.  Following physical
intuition~\cite{Seidl2007}, the optimal transport problem can be solved by
finding $N$ \emph{co-motion functions}
$\{\vf_{1}(\vr), \vf_{2}(\vr),\ldots,\vf_{N}(\vr)\}$, $\vf_i: \R^3 \to \R^3$. Each
$\vf_{i}(\vr)$ represents the 
optimal position of the $i$-th electron given the position of the first electron
at position $\vr$, with the natural definition that $\vf_{1}(\vr) = \vr$.  Since the electrons
are indistinguishable and distributed according to
the same density $\rho(\vr)$, the co-motion functions
should satisfy the mass conservation constraint that $\rho(\vf_{i}(\vr)) \ud\vf_{i}(\vr) =
\rho(\vr) \ud\vr$. Then $V_{ee}^{\SCE}[\rho]$ is given in terms of the co-motion
functions by~\cite{Seidl2007,OTformulation2012}
\begin{equation}
	V_{ee}^{\SCE}[\rho]	= \frac{1}{N} \int \rho(\vr)
	\sum_{i=1}^{N}\sum_{j>i}^{N} \frac{1}{\abs{\vf_{i}(\vr)-\vf_{j}(\vr)}} \ud\vr.
	\label{eqn:VeeComotion}
\end{equation}
The co-motion functions are implicit functionals of the electron
density, and can be obtained via semi-analytic formulations for
spherical symmetric atoms~\cite{Seidl2007,CFK2012} and strictly 1D
systems~\cite{MaletGiorgi2012}. However, these semi-analytic formulations are system
specific, and the co-motion functions
cannot be obtained in practice even for general
systems as simple as a dimer in 3D.

As an alternative to the co-motion framework, the Kantorovich
dual formulation of the optimal transport
problem~\cite{Kantorovich1942,Kantorovich2006,Villani2008,OTformulation2012,GangboSwiech1998}
introduces an auxiliary quantity called the Kantorovich potential
$u(\vr)$, in which
$V_{ee}^{\SCE}[\rho]$ can be obtained according to
\begin{multline}
	V_{ee}^{\SCE}[\rho] = \max_{u} \int u(\vs) \rho(\vs) \ud\vs, \\
	\text{s.t.~} \sum_{i=1}^{N} u(\vr_i) \le \sum_{i=1}^{N} \sum_{j>i}^{N}
	\frac{1}{\abs{\vr_i-\vr_j}},\quad\forall\, \{\vr_{i}\}_{i=1}^{N}.
	\label{eqn:Kantorovich}
\end{multline}
The Kantorovich dual problem~\eqref{eqn:Kantorovich} is a linear
programming problem with respect to $u$, and has the potential of
treating general systems with an arbitrary electron density.  However,
the Kantorovich problem introduces an infinite number of linear
constraints due to the arbitrary choice of $\{\vr_{i}\}_{i=1}^{N}$, and
cannot be directly implemented in practice.

Our novel method to overcome the difficulty of infinite number
of constraints in the Kantorovich problem reads as follows.
First note that the long-range asymptotic behavior
of the Kantorovich potential is
\begin{equation}
	u(\vr) = v(\vr) + C,
	\label{}
\end{equation}
where C is a constant chosen such that the function $v(\vr)$ vanishes at infinity
and satisfies~\cite{OTformulation2012,Seidl2007}
\begin{equation}
	v(\vr) \sim \frac{N-1}{\abs{\vr}} \quad\text{as}\quad \abs{\vr} \to \infty.
	\label{eqn:vAsymptotic}
\end{equation}
Without loss of generality  we refer to $v(\vr)$ also as the Kantorovich 
potential in the following discussion.
We introduce a functional $g[v]$ of $v(\vr)$ by
\begin{equation}
	g[v] = \min_{\{\vr_{i}\}} \sum_{i=1}^{N}\sum_{j>i}^{N} \frac{1}{\abs{\vr_i-\vr_j}}
	- \sum_{i=1}^{N} v(\vr_i), 
	\label{eqn:gfunctional}
\end{equation}
where the minimization is performed over all possible choices of the
positions of the $N$ electrons. The Kantorovich dual
problem~\eqref{eqn:Kantorovich} can then be written as
\begin{equation}
	\begin{split}
		V_{ee}^{\SCE}[\rho] = \max_{v,C} &\left(\int v(\vs)
		\rho(\vs) \ud\vs + N C\right), \\
	\text{s.t.~} &g[v]\ge N C.
	\end{split}
	\label{eqn:nestedKantorovich}
\end{equation}
Here we have used the normalization condition of
the electron density, $\int \rho(\vr) \ud\vr = N$.
Eq.~\eqref{eqn:nestedKantorovich} is a constrained optimization
problem with one inequality constraint, which can be solved by defining
a Lagrangian
\begin{equation}
	L[v, C, \lambda] = - \int v(\vs) \rho(\vs) \ud\vs - NC -
	\lambda \left( g[v] - NC \right).
	\label{eqn:Lagrangian}
\end{equation}
The Karush-Kuhn-Tucker (KKT) condition~\cite{NocedalWright1999} states 
that the optimal $(v^{*}(\vr),C^*,\lambda^{*})$ should satisfy the first
order necessary condition
\begin{subequations}
\begin{align}
	\frac{\delta L}{\delta v}(v^{*},C^*,\lambda^{*}) &= 0,\label{eqn:kkt1}\\
	\frac{\partial L}{\partial C}(v^{*},C^*,\lambda^{*}) & = 0,\label{eqn:kkt2}\\
	g[v^{*}] - NC^* & \ge 0,\label{eqn:kkt3}\\
	\lambda^{*}   & \ge 0,\label{eqn:kkt4}\\
	\lambda^{*} ( g[v^{*}] - NC^* ) & = 0.\label{eqn:kkt5}
\end{align}
\end{subequations}
Eq.~\eqref{eqn:kkt2} implies that
\begin{equation}
	\lambda^{*} = 1 > 0,
	\label{eqn:lambdastar}
\end{equation}
which satisfies Eq.~\eqref{eqn:kkt4}.  Combining Eq.~\eqref{eqn:lambdastar}
and \eqref{eqn:kkt5} (called the complementary slackness condition), we have
\begin{equation}
	g[v^{*}] = N C^{*}.
	\label{eqn:equalityConstraint}
\end{equation}
Therefore the constrained
optimization problem~\eqref{eqn:nestedKantorovich} can be solved by
eliminating the parameter $C$, resulting in a nested unconstrained
optimization problem (for
simplicity we drop the stars in the superscripts)
\begin{equation}
	V_{ee}^{\SCE}[\rho] = \max_{v} \left(\int v(\vs) \rho(\vs) \ud\vs
	+ g[v]\right).
	\label{eqn:equalityKantorovich}
\end{equation}
We also remark that
Eq.~\eqref{eqn:equalityKantorovich} can be viewed as a saddle point
problem 
\begin{equation}
	V_{ee}^{\SCE}[\rho] = \max_{v}\min_{\{\vr_i\}} h[v,\{\vr_i\}],
	\label{eqn:saddleKantorovich}
\end{equation}
with 
%\begin{equation}
%	h[v,\{\vr_i\}] =
%	\frac{1}{N}\Big(\int v(\vr) \rho(\vr) \ud\vr
%	+ \sum_{i=1}^{N}\sum_{j>i}^{N} \frac{1}{\abs{\vr_i-\vr_j}}
%	- \sum_{i=1}^{N} v(\vr_i)\Big).
%	\label{}
%\end{equation}
\begin{multline}
	h[v,\{\vr_i\}] =\\
	\int v(\vs) \rho(\vs) \ud\vs
	+ \sum_{i=1}^{N}\sum_{j>i}^{N} \frac{1}{\abs{\vr_i-\vr_j}}
	- \sum_{i=1}^{N} v(\vr_i).
	\label{}
\end{multline}
The numerical treatment of the saddle point
problem~\eqref{eqn:saddleKantorovich} is also difficult and is beyond the
scope of this paper.
%Since $v(\vr)$ has an infinite number of degrees of freedom and
%$\{\vr_i\}$ has $3 N$ degrees of freedom, the saddle point
%problem~\eqref{eqn:saddleKantorovich} finds a rank-$3 N$ saddle point
%of the functional $h[v,\{\vr_i\}]$.  To the extent of our knowledge,
%there is no efficient numerical method for finding such a saddle
%point directly in such a high dimensional space. 
%In practice 
Here we solve
Eq.~\eqref{eqn:equalityKantorovich} via a nested unconstrained
optimization approach.

The minimization problem~\eqref{eqn:gfunctional} for calculating $g[v]$ poses some hidden difficulties:
at any set of minimizers $\{\vr_i\}$, it must hold that
\begin{equation}
-\sum_{j \neq 1}^{N} \frac{\vr_1-\vr_j}{\abs{\vr_1-\vr_j}^3} - \nabla v(\vr_1) = 0.
\label{eqn:gfunctional_deriv}
\end{equation}
This is precisely Eq.~(7) in Ref.~\onlinecite{MaletGiorgi2012}.
Thus, at the exact dual potential $v(\vr)$, one recovers
the co-motion functions $\vf_i(\vr_1)$ by fixing $\vr_1$ and
minimizing~\eqref{eqn:gfunctional} with respect to $\vr_2, \dots, \vr_N$.
In particular, since~\eqref{eqn:gfunctional_deriv} holds for arbitrary
$\vr_1$, the minimizing set $\{\vr_i\}$ is not unique. As a consequence,
the functional derivative $\frac{\delta g[v]}{\delta v}(\vr)$ cannot be
analytically computed for the exact Kantorovich dual potential $v(\vr)$.
Thus we use derivative-free methods~\cite{Brent2002,NelderMeadSimplex1998} to
solve the outer optimization of the Kantorovich dual
problem~\eqref{eqn:equalityKantorovich}. The inner
optimization~\eqref{eqn:gfunctional} and~\eqref{eqn:gfunctional_deriv} for calculating $g[v]$ is a
standard optimization problem and is solved by the quasi-Newton method.
Our numerical results indicate that this hybrid approach can indeed
solve atoms and small molecules with reasonable parameterization of the
Kantorovich potential.

Special care should be taken when parameterizing $v(\vr)$ numerically.
The long range asymptotic behavior~\eqref{eqn:vAsymptotic} indicates
that the size of the computational domain needed to represent $v(\vr)$
is much larger than the size of the domain to represent the electron
density $\rho(\vr)$. For smooth $v(\vr)$, it turns out that the correct asymptotic
behavior of $v(\vr)$ can be efficiently preserved by introducing a \emph{pseudocharge}
associated to $v(\vr)$, denoted by
$m(\vr)$, i.e.,
\begin{equation}
	v(\vr) = \int \frac{m(\vr')}{\abs{\vr-\vr'}} \ud\vr'.
	\label{eqn:pseudocharge}
\end{equation}
The asymptotic behavior~\eqref{eqn:vAsymptotic} translates to the following
constraint on $m(\vr)$:
\begin{equation}
	\int m(\vr) \ud\vr = N-1.
	\label{eqn:normalizationM}
\end{equation}
Compared to $v(\vr)$ which decays as $(N-1)/\abs{\vr}$ for large $\abs{\vr}$,
physical intuition suggests that the support size of $m(\vr)$ should be much
smaller and is comparable to the support size for the electron density
$\rho(\vr)$, as shall be confirmed by our numerical results below.
Note that the parametrization in Eq.~\eqref{eqn:pseudocharge} does not
restrict the set of admissible $v(\vr)$ as long as $v(\vr)$ is sufficiently
smooth, since one can simply define $m(\vr)$ as $- \Delta v/(4\pi)$.

For strictly 1D systems however, we remark that the co-motion functions are
discontinuous~\cite{CFK2012,OTformulation2012}. In particular, Eq.~\eqref{eqn:gfunctional_deriv}
implies that $\nabla v(\vr)$ is discontinuous for these systems, and the
pseudocharge will consist of $\delta$-functions and is difficult to
discretize. For strictly 1D systems, we therefore discretize $v(\vr)$ directly
on a grid which matches the asymptotic condition~\eqref{eqn:vAsymptotic}.

\section{Numerical results}\label{sec:numerical}

\paragraph{Beryllium atom.}
To illustrate the performance of the nested optimization method in
practice, we first study the beryllium atom with $4$ electrons. 
Similar to Ref.~\onlinecite{Seidl2007}, the electron density is provided
non-self-consistently by a configuration interaction calculation with
Slater-type orbitals~\cite{NuclearChargeLimit2009,AsymptoticsCI2009}.
Specifically, $\rho(\vr)$ is a linear combination of terms $r^j
\mathrm{e}^{-\lambda\,r}$ with $j = 0, 1, 2$.  Since $\rho(\vr)$ for
beryllium is spherically symmetric, the co-motion functions can be
obtained semi-analytically~\cite{Seidl2007} with numerical optimization
performed on the angular part of each co-motion function. Our calculation gives
$V_{ee}^\mathrm{SCE}[\rho] = 4 \times 0.812132$.

\begin{figure}[!ht]
	\centering
	\includegraphics[width=0.7\columnwidth]{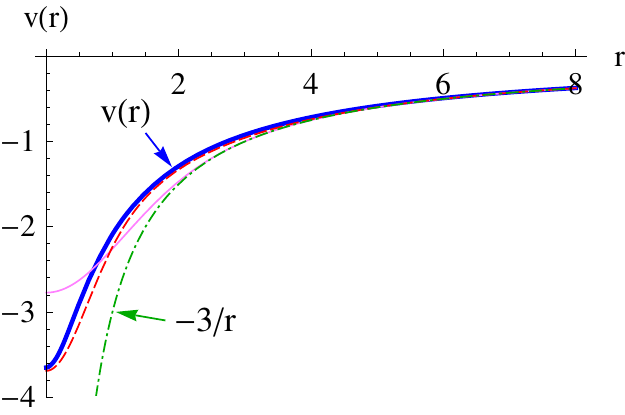}
	\caption{(Color online) Kantorovich potential $v(\vr)$ for the beryllium
	atom: co-motion formulation (thick blue solid line), Kantorovich dual
	formulation with the pseudocharge $m(\vr)$ parametrized by a single
	Gaussian (thin magenta solid line) and by the sum of two Gaussians (red
	dashed line).  The green dot-dashed line shows the asymptotic
	expansion~\eqref{eqn:vAsymptotic}.} \label{fig:v_OT_Be}
\end{figure}

For the Kantorovich dual formulation, we try to expand the pseudocharge
$m(\vr)$ as a linear combination of Gaussian basis functions.
For simplicity, we initially parametrize the
pseudocharge $m(\vr)$ by a single Gaussian function as
\begin{equation}
	m(\vr;\sigma) = \frac{N-1}{\left(2\pi \sigma^2\right)^{3/2}}\,
	\mathrm{e}^{-\frac{\vr^2}{2\sigma^2}},
	\label{eqn:pseudochargeGauss}
\end{equation}
with the value $\sigma$ left to be determined in the optimization
procedure.  The corresponding Kantorovich potential has the
analytic form
\begin{equation}
	v(\vr;\sigma) = \int \frac{m(\vr')}{\abs{\vr-\vr'}} \ud \vr'
	= \frac{N-1}{\abs{\vr}}\, \mathrm{erf}\!\left(\frac{\abs{\vr}}{\sqrt{2}\,\sigma}\right).
	\label{eqn:vGauss}
\end{equation}
The
nested optimization method gives $V_{ee}^{\SCE,1}[\rho] = 4 \times
0.647$ with $\sigma = 0.8630$, and the relative error of
$V_{ee}^\mathrm{SCE}$ is $20.3\%$.  The result can be significantly
improved by parameterizing the pseudocharge $m$ by a sum of two
concentric Gaussian functions:
\begin{equation}
	m(\vr) = (N-1) \left( \cos^2(\vartheta) \frac{\mathrm{e}^{-\frac{\vr^2}{2\sigma_1^2}}}
	{\left(2\pi \sigma_1^2\right)^{3/2}} +
	\sin^2(\vartheta) \frac{\mathrm{e}^{-\frac{\vr^2}{2\sigma_2^2}}}
	{\left(2\pi \sigma_2^2\right)^{3/2}} \right),
	\label{eqn:pseudochargeTwoGauss}
\end{equation}
which yields $V_{ee}^{\SCE,2}[\rho] = 4 \times 0.7995$ with
parameters $\sigma_1 = 0.4507,\ \sigma_2 = 1.862,\ \vartheta = 0.6872$.
The relative error of $V_{ee}^{\SCE}[\rho]$ is significantly reduced to
$1.6\%$, which is quite small given that only $3$ parameters are
employed. The
corresponding Kantorovich potential $v(\vr)$ is shown in
Fig.~\ref{fig:v_OT_Be}, in comparison to the (numerically) exact
potential obtained via the co-motion formulation. As for
$V_{ee}^\mathrm{SCE}$, the potential $v(\vr)$ with
pseudocharge~\eqref{eqn:pseudochargeTwoGauss} agrees remarkably well
with the exact potential. To further improve the result, one can use
a larger number of Gaussian basis functions to represent $m(\vr)$.
Note that this Ansatz implicitly assumes that the exact $v(\vr)$ is smooth,
which is indeed the case for this example.

\paragraph{Quantum wire.}
Next we study a model quantum wire system in 1D, for which the co-motion
formulation can also be solved semi-analytically~\cite{MaletGiorgi2012}. 
The system consists of $N = 4$ electrons and the Hamiltonian reads
\begin{equation}
	H = -\frac{1}{2} \sum_{i=1}^N \frac{\partial^2}{\partial x_i^2}
	+ \sum_{i=1}^{N} \sum_{j>i}^N w_b(x_i - x_j)
	+ \sum_{i=1}^N v_{\ext}(x_i),
\end{equation}
where $v_{\ext}(x) = \frac12 \omega^2 x^2$ is a confining
potential and
\begin{equation}
w_b(x) = \frac{\sqrt{\pi}}{2\,b}\,
\mathrm{exp}\!\left(\frac{\abs{x}^2}{4\,b^2}\right)
\mathrm{erfc}\!\left(\frac{\abs{x}}{2\,b}\right)
\end{equation}
is the effective Coulomb
interaction. By increasing the length scale $L \equiv 2 \omega^{-1/2}$,
the system approaches the SCE limit due to the long-range effective
Coulomb interaction $w_{b}(x)$. Concretely, as $L$ increases
from $4.5$ to $14$, the quantum wire system
transforms from a weakly correlated system with $2$ peaks in the
electron density to a strongly correlated system with $4$ peaks in the
electron density~\cite{MaletGiorgi2012}, which cannot be described by
the local density approximation (LDA)~\cite{PerdewZunger1981} of the KS
exchange-correlation functional.

We discretize the Hamiltonian by Hermite functions.  The electron
density is represented numerically on a grid and is obtained via
self-consistent field iterations (SCF). 
In the Kantorovich dual formulation, we avoid the pseudocharge formulation
for this example since the derivative $v'(x)$ of the exact dual
potential is not continuous~\cite{CFK2012,OTformulation2012} and the pseudocharge consists of
$\delta$-functions. Instead, we discretize $v(x)$ directly on a
uniform grid. The number of grid points is a compromise
between accurate parametrization of $v(x)$ and feasibility of the
optimization~\eqref{eqn:equalityKantorovich}.
We focus on the cases $L = 6$ and $L = 14$, and
choose the grid spacing $\Delta x_L$ somewhat heuristically as
$\Delta x_6 = \tfrac{3}{2}$ and $\Delta x_{14} = 4$.
We allow $v(x)$ at the grid points $-M_L, -M_L + \Delta x_L, \dots, M_L$ with
$M_6 = 7.5$ and $M_{14} = 28$ to be determined by the optimization
procedure, and fix $v(x)$ by the asymptotic formula~\eqref{eqn:vAsymptotic}
at grid points $\abs{x} > M_L$.
Between grid points, we use piecewise cubic Hermite interpolation.
Additionally, due to the even symmetry $v(x) = v(-x)$ it suffices to
optimize $v(x)$ for $x \ge 0$ only.
The interval $[-M_L, M_L]$ (almost) covers the support of the electron density $\rho(x)$
and corresponds to the characteristic shape of $\rho(x)$,
which we try to reproduce by the SCF iteration. 
As starting point for the SCF iteration,
we convolve the exact $v(x)$ and $\rho(x)$ from the co-motion formulation
with a Gaussian with variance $\tfrac{L}{6}$ and $\tfrac{L}{4}$, respectively.
We use linear mixing with parameter $\lambda = 0.1$.

\begin{figure}[!ht]
\centering
\subfloat[]{\includegraphics[width=0.48\columnwidth]{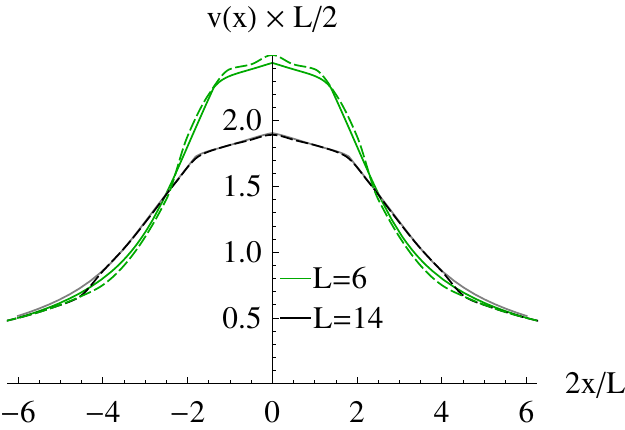}}
\hspace{0.02\columnwidth}
\subfloat[]{\includegraphics[width=0.48\columnwidth]{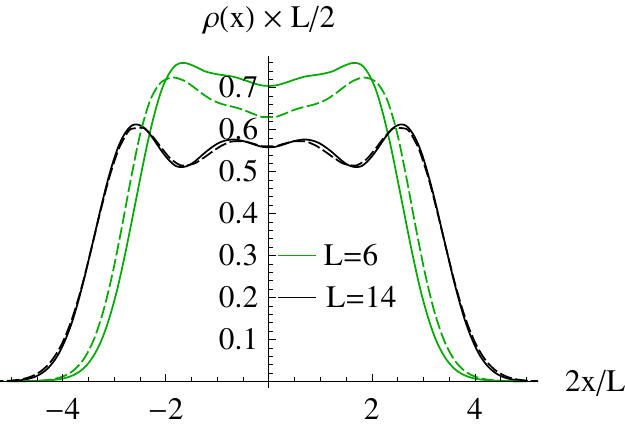}}
\caption{(Color online) Comparison of the Kantorovich potential $v(x)$ (a) and
density $\rho(x)$ (b) obtained via the ``exact'' co-motion formulation~\cite{MaletGiorgi2012}
(solid lines) and our dual formulation (dashed lines), respectively.
The green (upper) curves correspond to $L = 6$,
and the black (lower) curves correspond to $L = 14$.}
\label{fig:v_rho_quantum_wire}
\end{figure}

Fig.~\ref{fig:v_rho_quantum_wire} shows the Kantorovich
potential $v(x)$ and the density $\rho(x)$ obtained via our method (after $15$ and $25$
SCF iterations for $L = 6$ and $L = 14$, respectively), in comparison to the
(numerically exact) co-motion formulation.  The Kantorovich dual
formulation correctly reproduces the strong interaction limit ($L=14$)
with $4$ peaks in the electron density.
While the results match quite well for $L = 14$, one notices a deviation
of the density $\rho(x)$ from the co-motion reference for $L = 6$. This
observation is also reflected by the values of $V_{ee}^{\SCE}$ (after
the SCF iteration)
shown in Table~\ref{tab:KdualQuantumWire}.
\begin{table}[!ht]
\begin{tabular}{r|cc}
L &6 &14\\
\hline
``exact'' $V_{ee}^{\SCE}$& 1.025& 0.3408\\
\hline
dual-K $V_{ee}^{\SCE}$& 0.9394& 0.3381\\
relative error&  8.4\%& 0.8\% \\
\end{tabular}
\caption{$V_{ee}^{\SCE}$ of the model quantum wire system in 1D,
for the co-motion formulation (reference) and
the Kantorovich dual formulation.}
\label{tab:KdualQuantumWire}
\end{table}
Namely, the relative error of $V_{ee}^{\SCE}$ for $L = 14$ is much
smaller than for $L = 6$. The deviation is likely due to numerical
difficulties in the maximization~\eqref{eqn:equalityKantorovich}.  As
mentioned above, we make use of the Nelder-Mead simplex
algorithm~\cite{NelderMeadSimplex1998} which is a derivative-free
optimization method for the outer optimization.  Unfortunately, the
results shown in Fig.~\ref{fig:v_rho_quantum_wire} depend quite
sensitively on the parametrization of $v(x)$, e.g., the choices of the
above $\Delta x_L$ and $M_L$.  For different choices, $v(x)$ might
acquire local maxima during the SCF iteration. Thus further
improvements of the optimization~\eqref{eqn:equalityKantorovich} are
required, which we leave as work for the future.

\paragraph{Trimer molecule.}

\begin{figure}[htpb]
\centering
\subfloat[]{\includegraphics[width=0.4\columnwidth]{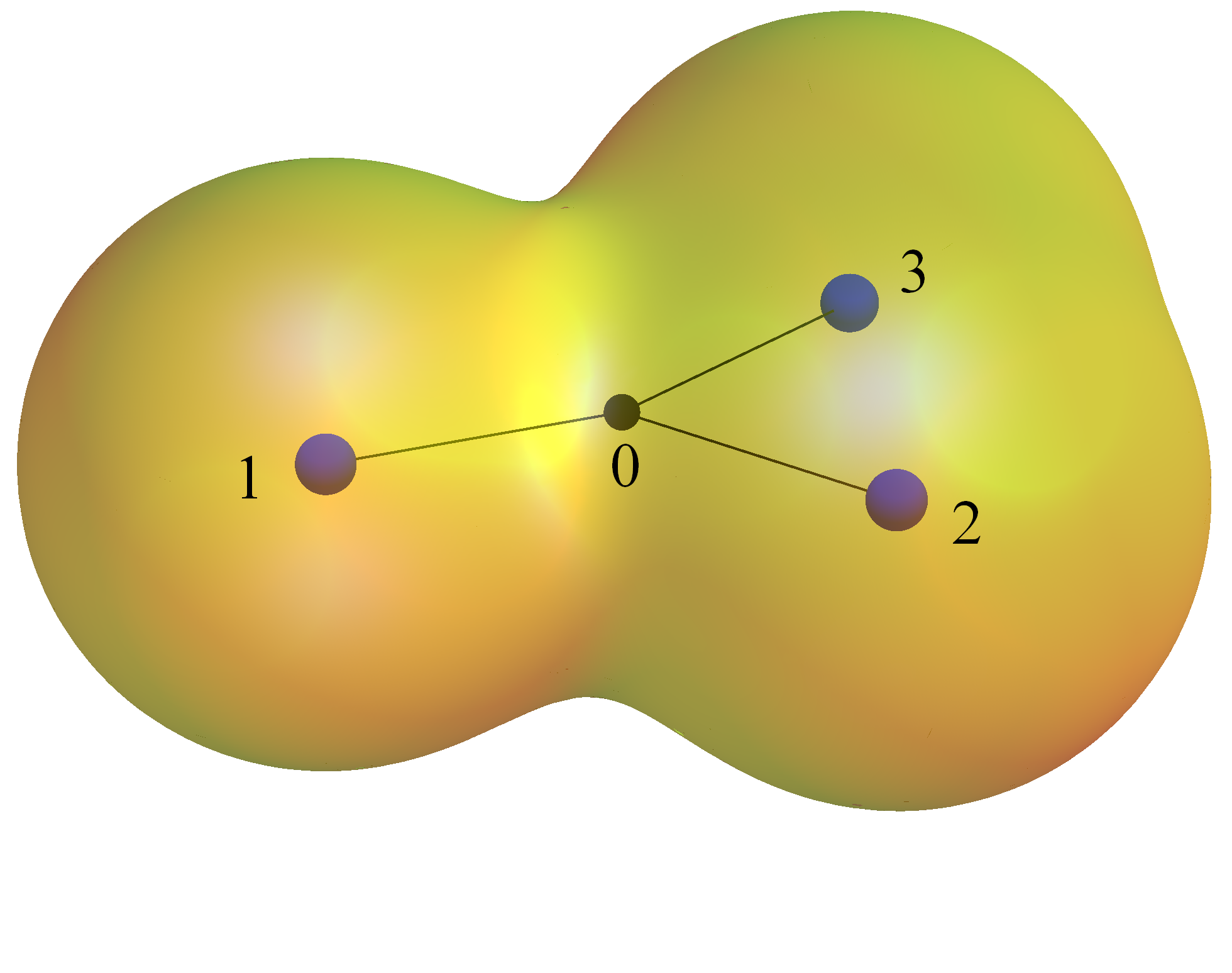}
\label{fig:m_OT_trimer_iso}}
\subfloat[]{\includegraphics[width=0.6\columnwidth]{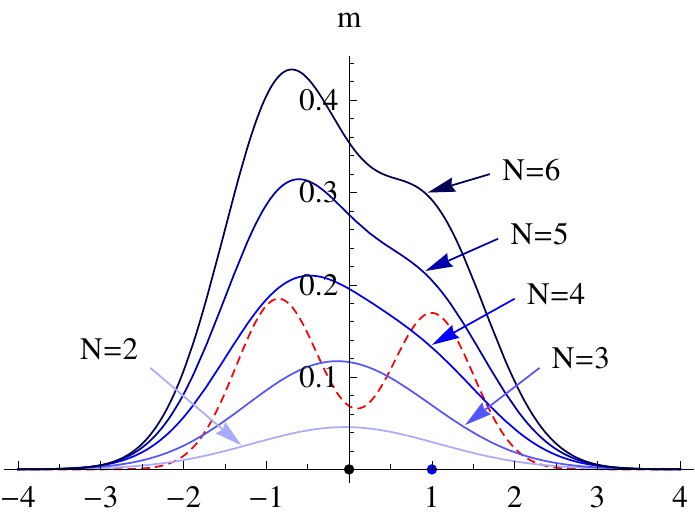}
\label{fig:m_OT_trimer_pseudocharge}}
\caption{(Color online) (a) An %The $\tfrac{N}{42}$-
isosurface of the
electron density $\rho(\vr)$ (normalized to $1$) of a model trimer.
(b) Optimized pseudocharge $m$ of the trimer molecule (solid blue)
and density $\rho(\vr)$ as is in (a) (dashed red),
plotted along the line connecting $1$ and $0$ in (a).
The values $0, 1$ on the x-axis match the corresponding points in (a).}
\label{fig:m_OT_trimer}
\end{figure}

Finally we apply our method to a model trimer in 3D, for which the
optimal transport problem cannot be solved with known techniques using
the co-motion formulation. For simplicity, the electron density
$\rho(\vr)$ is given non-self-consistently by a sum of three Gaussian
functions centered at the points $1, 2, 3$ in Fig.~\ref{fig:m_OT_trimer_iso},
respectively. Each Gaussian has variance
$\frac{1}{2}$, and each point $1, 2, 3$ has distance $1$ from the
origin. The normalization of $\rho(\vr)$ is fixed by the number of
electrons $N = 2,3,4,5,6$. An isosurface of the electron density
is shown in yellow in Fig.~\ref{fig:m_OT_trimer_iso}.

We parametrize the pseudocharge (similar to the density)
by a sum of $3$ Gaussian functions with the same variance
$\sigma$.  The centers of the Gaussian functions are located on the
black lines in Fig.~\ref{fig:m_OT_trimer_iso} connecting $0 \to
1$, $0 \to 2$ and $0 \to 3$ respectively, with equal
distance $R$ from the origin. The variance $\sigma$ and the distance
$R$ are to be determined by the optimization procedure. The results are
summarized in Table~\ref{tab:trimer}, including $V_{ee}^{\SCE}$.
Unlike the previous cases for the Be atom and the one dimensional
system, to the extent of our knowledge there is no available method that
allows us for benchmarking the accuracy of $V_{ee}^{\SCE}$ for the
trimer molecule.  Nevertheless, the values in Table~\ref{tab:trimer} are
lower bounds on $V_{ee}^{\SCE}$ due to the Ansatz for $m(\vr)$
in the maximization~\eqref{eqn:equalityKantorovich}.

\begin{table}[!ht]
\begin{tabular}{c|c@{\hspace{6pt}}c@{\hspace{6pt}}c@{\hspace{6pt}}c@{\hspace{6pt}}c}
$N$& 2& 3& 4& 5& 6\\
\hline
$V_{ee}^{\SCE}$& 0.1973& 0.4617& 0.7584& 1.0711& 1.3959\\
$\sigma$& 1.1073& 0.9804& 0.8313& 0.7741& 0.7315\\
$R$&      0.2260& 0.5322& 0.8538& 0.9062& 0.9417\\
\end{tabular}
\caption{$V_{ee}^{\SCE}$ of the trimer molecule,
and corresponding optimized pseudocharge parameters $\sigma$ and $R$.
Due to the specific Ansatz for the pseudocharge (described in the text),
the values for $V_{ee}^{\SCE}$ should be regarded as lower bounds.}
\label{tab:trimer}
\end{table}
Fig.~\ref{fig:m_OT_trimer_pseudocharge}
shows the optimized pseudocharge $m(\vr)$ plotted along the line $0
\rightarrow 1$ as in Fig.~\ref{fig:m_OT_trimer_iso}.  Along with
increasing $N$, the magnitude of the pseudocharge increases as required
by the normalization condition~\eqref{eqn:normalizationM}. The shape of
the pseudocharge develops from a unimodal function for $N = 2$ to a
bimodal function for $N = 6$, indicating growing influence of the distance
$R$. For large $N$ the bimodal pseudocharge is biased towards the negative axis
around atoms $2$ and $3$ where the electron density is larger. The
increase of $R$ is accompanied by a decrease of $\sigma$, and the
support size of the pseudocharge remains approximately the same as $N$
increases, and is comparable to the support size of $\rho(\vr)$.

\section{Conclusion and future work}\label{sec:conclusion}

In this paper we present a nested optimization method for solving the
Kantorovich dual problem to obtain the exchange correlation functional
in the SCE limit for strongly correlated systems.  With reasonable
parameterization which preserves the asymptotic property of the
Kantorovich potential, the Kantorovich dual solution can be obtained for
atoms and small molecules.  Based on the Kantorovich dual formulation,
one can combine the SCE exchange-correlation functional with existing
exchange-correlation functionals for Kohn-Sham non-interacting
kinetic energy in order to improve the performance of KSDFT for
strongly correlated systems. For instance, we may mix the SCE
exchange-correlation functional with the GGA exchange-correlation
functional~\cite{Becke1988,LeeYangParr1988,PerdewBurkeErnzerhof1996} via
linear combination as
\begin{equation}
	E^{\mathrm{SCE-GGA}}[\rho] = (1-\alpha) E^{\mathrm{SCE}}[\rho] + \alpha
E^{\mathrm{GGA}}[\rho],
	\label{}
\end{equation}
and obtain $\alpha \in [0,1]$ via a set of benchmark problems.

Due to the difficulty in obtaining the functional derivative
$\frac{\delta g[v]}{\delta v}(\vr)$ by an analytic formula, in practice
the outer optimization of the nested optimization method is solved by
derivative-free optimization methods. However, our numerical results
indicate that the derivative-free methods may get stuck at local minima.
Moreover, the derivative-free methods are not suitable for optimizing
with respect to a large number of degrees of freedom.  More efficient
numerical methods need to be developed in order to obtain the
Kantorovich dual solution for more general systems in the future.

This work is partially supported by the Technische Universit\"at
M\"unchen (C.~M.), and the Laboratory Directed Research and Development
Program of Lawrence Berkeley National Laboratory under the U.S.
Department of Energy contract number DE-AC02-05CH11231 (L.~L.).  C.~M.
thanks the hospitality of the Lawrence Berkeley National Laboratory
where the idea of this work starts.  We thank Codina Cotar, Gero
Friesecke and Brendan Pass for many helpful discussions, as well as
Paola Gori-Giorgi for sharing numerical details concerning the quantum
wire model.

\bibliographystyle{apsrev}
%\bibliography{references}

\end{document}